# INFLUENCE OF EGR COMPOUNDS ON THE OXIDATION OF AN HCCI-DIESEL SURROGATE

J.M. ANDERLOHR[*1,3], A. PIPEREL[1,2], A. PIRES DA CRUZ[1],

R. BOUNACEUR[3], F. BATTIN-LECLERC[3], P. DAGAUT[2], X. MONTAGNE[1]

[1]Institut Français du Pétrole,

1 et 4, Ave. Bois Préau, 92852 Rueil Malmaison Cedex, France

[2]CNRS

1C, Ave. de la Recherche Scientifique, 45071 Orléans Cedex, France

[3]Département de Chimie-Physique des Réactions, Nancy Université, CNRS,

ENSIC, 1, rue Grandville, BP 20451, 54001 Nancy Cedex, France

---

[*] E-mail : Jorg.ANDERLOHR@ifp.fr; Tel.: +33 1 47 52 60 00

**ABSTRACT**

This paper presents an experimental and numerical study of the impact of various additives on the oxidation of a typical automotive surrogate fuel blend, i.e. n-heptane and toluene. It examines the impact of engine re-cycled exhaust gas compounds on the control of a Homogeneous Charge Compression Ignition (HCCI) engine. A series of experiments were performed in a highly diluted Jet-Stirred Reactor (JSR) at pressures of 1 and 10atm (1atm = 101325Pa). The chosen thermo-chemical conditions were close to those characteristic of the pre-ignition period in an HCCI-engine. The influence of various additives, namely nitric oxide (NO), ethylene ($C_2H_4$) and methanol ($CH_3OH$), on the oxidation of a n-heptane/toluene blend was studied over a wide range of temperatures (550-1100K), including the zone of the Negative Temperature Coefficient (NTC).

A new detailed chemical kinetic reaction mechanism is proposed for the oxidation of the surrogate fuel. It includes reactions of NO, methanol ($CH_3OH$) and ethylene ($C_2H_4$) and is used to explain the obtained experimental data. The mechanism is further used to theoretically study the impact of other Exhaust Gas Recirculation (EGR) compounds on the hydrocarbon oxidation, namely ethane ($C_2H_6$), formaldehyde (HCHO) and carbon monoxide (CO).

**Keywords :** HCCI, EGR, JSR, kinetic modelling, n-heptane/toluene blend.

# 1 INTRODUCTION

The mode of Homogeneous-Charge-Compression-Ignition (HCCI) in internal combustion engines is currently the subject of many research projects [1-3]. The HCCI-engine combines advantages of Spark Ignition (SI) engines with those of Compression Ignition (CI) engines. The homogenous fuel/air mixture guarantees low particulate emissions and the high dilution permits very low production of nitric oxides ($NO_x$), while the principle of CI assures a high efficiency close to that of a diesel engine. The main challenge of operating HCCI combustion engines is the control of auto-ignition. Various strategies for auto-ignition control are known [1]. A promising one is Exhaust Gas Recirculation (EGR) [4]. The principle of EGR is to induct an exhaust gas fraction into the combustion chamber together with the fresh gases. EGR has also the advantage of reducing the maximum combustion temperature in the engine and so the production rate of thermal-NO. However, HCCI exhaust gases are composed of many different species including products from condensed fuel, Unburned HydroCarbons (UHC) and traces of $NO_x$. Piperel et al. [8] have recorded along the EGR pipe concentrations of CO between 0.032 and 0.74%, NOx between 2 and 100 ppmv, ethylene between 45 and 115 ppmv and formaldehyde between 20 and 50 ppmv. It is known that the dilution by EGR impacts strongly on HCCI-engine control, but it is also supposed that the chemical composition of the EGR-gases governs HCCI combustion processes. While the impact of recycled $NO_x$ is known to be important for the kinetics of oxidation of UHC [5-7], the influence of other recycled species has not yet been comprehensively assessed [4]. A n-heptane/toluene mixture was used as a diesel surrogate with n-heptane having a cetane number close to diesel and toluene representing the aromatic fraction of a diesel fuel.

In the first part of this study, we focus on the impact of nitric oxide (NO), ethylene ($C_2H_4$) and methanol ($CH_3OH$) on the surrogate fuel oxidation. Ethylene is an important intermediate in the combustion of hydrocarbons and therefore it is present in the EGR. Methanol is a potential component of future automotive fuel blends [9] and the comprehension of its interactions with the proposed automotive surrogate n-heptane/toluene is of interest.

A series of experiments were performed to evaluate the influence of NO, ethylene and methanol on the oxidation of the surrogate fuel using a Jet Stirred Reactor (JSR) over a wide range of temperatures, covering the zone of the Negative Temperature Coefficient (NTC). The experiments were limited to fuel-lean conditions (equivalence ratio 0.75), relevant for HCCI operation, and were conducted at 1 and 10atm (1atm = 101325Pa).

A detailed chemical kinetic reaction mechanism for the oxidation of a n-heptane/toluene mixture including effects of NO, ethylene and methanol addition is proposed to explain the experimental data obtained. In the second part of this study, the mechanism was used to numerically analyze the impact of other EGR components on the fuel oxidation, namely ethane ($C_2H_6$), formaldehyde (HCHO) and carbon monoxide (CO).

## 2    EXPERIMENTAL PROCEDURE AND RESULTS

The experiments were performed in a fused silica JSR heated by a regulated electrical resistance system. The JSR and heating system are surrounded by insulating material and a pressure-resistant jacket, allowing operating pressures up to 10atm [6]-[9]. It consists of a sphere with an internal volume of 29.5cm$^3$ equipped with four injectors. N-heptane (99% pure) and toluene (99% pure) were premixed at the desired composition after thorough ultrasonic degassing. The fuel was pumped using a micro piston High-Performance-Liquid-Chromatography (HPLC) pump and sent to an atomizer-vaporizer assembly. Thermal mass-flow controllers were used for all gases. For atomization, a nitrogen gas-flow of 0.05m$^3$/h was used. Samples were taken at steady state using a sonic quartz probe.

A high degree of dilution was applied (~98% in mole of molecular nitrogen) minimizing the fuels' heat release. Temperature gradients measured inside the reactor by a thermocouple lay below 10K (typically 2-5K) while concentration gradients could not be detected. Good thermal and mixture homogeneities, as well as experimental reproducibility, equivalent to those obtained in previous experiments performed with the same equipment were achieved [10], [11]. Samples of reacting mixtures were taken by sonic probe sampling and collected in $1 \times 10^{-3} m^3$ Pyrex bulbs at around $40 \times 10^{-3}$ atm for immediate analysis with two Gas-Chromatographs (GC). One GC was equipped with a Flame-Ionization-Detector (FID) and the other one with a Thermal-Conductivity-Detector (TCD). In addition, on-line Fourier-Transform-InfraRed (FTIR) analyses were performed. The sampling probe was connected to a temperature controlled gas cell (413K) via a heated Teflon tube. The sample pressure in the cell was 0.2atm. This equipment enabled the measurements of water ($H_2O$), carbon dioxide ($CO_2$), CO, $C_1$-$C_5$ hydrocarbons, n-heptane and toluene (by FTIR-GC) and permanent gases, molecular hydrogen ($H_2$) and oxygen ($O_2$) (by TCD-GC). The carbon balance was checked for every sample (100±5%). In order to observe the fuel oxidation from low to high conversion, the experiments were performed at a constant mean residence time of 0.7s at 10atm (0.07s at atmospheric pressure) and an equivalence ratio of 0.75 by increasing stepwise the gas-temperature inside the reactor from 560 to 1190K.

Experiments were performed for the oxidation of the n-heptane/toluene blend with addition of various concentrations of methanol, ethylene and NO. Because at atmospheric pressure the addition of these species shows only little impact on the oxidation of n-heptane and toluene due to a lower global reactivity, results are only shown for a pressure of 10atm. Figure 1 presents the measured concentrations of different species as functions of temperature at an equivalence ratio of 0.75 for pure fuel, a fuel/NO mixture and a mixture of fuel/ethylene/methanol. The initial concentrations of n-heptane and toluene were held constant at 900 and 100 ppmv respectively, and the influence of the additives methanol (130ppmv), ethylene (200ppmv) and NO (100ppmv) was tested separately and together. The additive concentrations chosen may slightly exceed those found in EGR [8], but the objective is to understand their general impact on fuel oxidation. The impact of the addition of ethylene on the mixture's reactivity as well as on the formation of intermediates is similar to that caused by the addition of methanol. Experimental results are then only shown for the addition of both ethylene and methanol together. Concerning the addition of NO, results are shown for mixtures of n-heptane/toluene with the addition of NO only.

**FIGURE 1**

Figures 1a (n-heptane) and 1b (toluene) show that the addition of NO inhibits the oxidation of the fuel at low temperatures (<700K), accelerates it at high temperatures (>850K) and reduces the intensity of the NTC. The accelerating effect above 800K is also observed on the oxygen-concentration profile (Fig.1c), showing that the addition of NO induces a strong depletion above 900K. This result agrees with the observations of Dubreuil et al. [6] and Moréac et al. [7] in their studies on the complex impact of NO on HC oxidation. At temperatures below 850K, the addition of ethylene and methanol leads to a slightly retarded oxidation of the fuel, whereas at higher temperatures, this addition has no impact on the fuel oxidation.

For temperatures below 850K, the effect of NO on the ethylene concentration profile (Fig.1d) is negligible. Above 850K, the formation of ethylene is strongly reduced when NO is added and the ethylene peak-concentration, observed at 850K, decreases strongly. Below 850K, when ethylene is added, its measured concentrations are increased by the magnitude of the initially added amounts. At higher temperatures, ethylene addition does not affect its measured concentration profiles, which become quasi independent of the initially added ethylene concentration.

Figure 1e shows that generally, at around 650K, a production peak of methanol from the fuel oxidation is observed. This production is reduced when NO is added. For temperatures above 700K and in the case of methanol-addition, the methanol concentration profile includes also a NTC-zone. One observes that the ethane concentration (Fig.1f) is not greatly affected by the addition of ethylene and methanol, but is strongly reduced when NO is added. The presence of NO leads to an increase of the formaldehyde concentration in the NTC-region (700-800K) and to a decrease at higher temperatures. This result is consistent with the observations made by Dagaut et al. [12]. At temperatures below 850K the addition of methanol and ethylene increases the measured formaldehyde concentration (Fig.1g), but at higher temperatures, the formaldehyde concentration profile is unchanged by their addition.

Ethylene and methanol addition has little effect on the formation of CO (Fig.1h), $H_2$ (Fig.1i), $H_2O$ (Fig.1j) and $CO_2$ (Fig.1k). At temperatures above 800K, these additives slightly increase the concentrations of these oxidation products. In contrast, the addition of NO increases the CO and hydrogen production between 800K and 900K, and considerably decreases it at higher temperatures. Consequently, the formation of $H_2O$ and $CO_2$ is also increased by the addition of NO at these temperatures.

## 3    DESCRIPTION OF THE DETAILED KINETIC MODEL

To explain the obtained experimental results, a new kinetic reaction mechanism was developed and used to simulate the oxidation of n-heptane/toluene blends in presence of ethylene, methanol and $NO_x$. This model is based on a mechanism generated by <u>E</u>XGAS-<u>A</u>LKANES (EA) according to the rules described by Buda et al. [13] and on the toluene mechanism of Bounaceur et al. [14]. It also includes reactions for the co-oxidation of n-heptane and toluene. Thermochemical data for molecules and radicals were calculated by the THERGAS software [15], which is based on group and bond additivity methods proposed by Benson [16] and yields 14 polynomial coefficients according to the CHEMKIN-II formalism [17].

### 3.1    Mechanism for the oxidation of n-heptane

The mechanisms generated by EA include three sub sets [13], [18]:

➢ A $C_0$-$C_2$ reaction base [19] involving species with up to two carbon atoms and including kinetic data mainly taken from the evaluations of Baulch et al. [20] and Tsang and Hampson [21]. The reactions of ethylene and methanol are part of this base. Reactions of methanol are mainly taken from Tsang [22]. The pressure dependent rate coefficients are obtained using the formalism proposed by Troe [23].

➢ A comprehensive primary mechanism. It considers only initial organic compounds together with oxygen as reactants. It includes all the usual low and intermediate temperature reactions of alkanes (unimolecular and bimolecular initiations, additions to oxygen, radical isomerizations, formation of conjugated alkenes, smaller alkenes, cyclic ethers, aldehydes, ketones, hydroperoxides species, H-abstractions by small radicals, disproportionation between alkylperoxy and hydroperoxy (•OOH) radicals). The disproportionation between methylperoxy and other alkylperoxy radicals, leading to the formation of oxygen, aldehyde and methanol molecules, as proposed by Lightfoot et al. [24], was also added in order to explain the observed formation of methanol around 650K.

➢ A lumped secondary mechanism [18]. The molecules formed in the primary mechanism, with the same molecular formula and the same functional groups, are lumped into one unique species without distinction between the different isomers. It includes global reactions producing, in the smallest number of steps, molecules or radicals whose reactions are included in the $C_0$-$C_2$ base. These reactions produce $C_1$-$C_2$ species in large amounts. The standard EA [13] mechanisms tend to over-predict ethylene concentrations at low temperatures. Since the modeling of ethylene concentrations is of major importance for this study, the ethylene producing reactions, defined in the secondary mechanism, were improved. The applied modifications concern the decompositions of hydroperoxides and cyclic ethers, as well as the H-abstractions by •O•, hydroxy (•OH), •OOH and methyl radicals (•$CH_3$) from olefins. This promotes the formation of larger alkyl or alkenyl radicals, instead of producing small $C_0$-$C_2$-radicals in one step reaction. The improved mechanism will be further referenced as modified EA-version.

*3.2  Mechanism for the oxidation of toluene and co-oxidation reactions*

The toluene model includes sub-mechanisms for :

➢ The oxidation of benzene.

➢ The oxidation of unsaturated $C_3$-$C_4$ species.

➢ A primary mechanism including reactions of toluene and benzyl, methylphenyl, peroxybenzyl, alcoxybenzyl and cresoxy free radicals.

> A secondary mechanism involving reactions of benzaldehyde, benzyl hydroperoxyde, cresol, benzylalcohol, ethylbenzene, styrene and bibenzyl [14].

Due to the strong electronic delocalization of benzyl radicals, the following co-oxidation reactions with alkyls are considered for the oxidation of toluene [25]:

- Metathesis of benzyl radicals with n-heptane yielding toluene and heptyl radicals.
- Terminations between benzyl and heptyl radicals.
- Metathesis of secondary allylic radicals with toluene.
- Reactions between toluene and alkylperoxy radicals.

*3.3 Mechanism of nitrogen containing species*

Reactions of nitrogen containing compounds were taken from a previous modeling work on the conversion of NO to nitric dioxide ($NO_2$) promoted by methane, ethane, ethylene, propane, and propene [26]. The kinetic model was primarily based on the GRI-MECH-3.0 [27] and the research performed by Dean and Bozzelli [28] and Atkinson et al. [29]. The thermochemical data proposed by Marinov [30] was used for nitrogen compounds. N-heptane-$NO_x$ reactions and rate constants were derived from a mechanism published by Glaude et al. [31] for the oxidation of n-butane and n-pentane in the presence of $NO_x$. These include:

> Reactions of alkylperoxy ROO• and hydroperoxy-alkylperoxy (HOOQOO•) radicals with NO leading to $NO_2$ and partially oxidized products. ROO• radicals react with NO forming $NO_2$ and alkoxy radicals RO•. Furthermore, the decomposition of RO• radicals was included. HOOQOO• radicals reacting with NO are decomposed by a global reaction to $NO_2$, a hydroxyl radical, two formaldehyde molecules and the corresponding alkene.

> Reactions of resonance stabilized allylic radicals with $NO_2$ yielding NO, acrolein and an alkyl radical.

Toluene-$NO_x$ reactions are also proposed within the framework of the present study. These include:

> Conversion of benzyl with $NO_2$ generating alkoxybenzyl radicals with the same reaction-rate constants as defined for the reaction between methyl radicals and $NO_2$ [32].

- Reactions of peroxybenzyl and peroxyphenyl radicals with NO yielding alkoxybenzyl and phenoxy radicals with the same reaction-rate constants as defined for alkylperoxy radicals [31].
- Reactions of alkoxybenzyl radicals with NO and $NO_2$ yielding benzaldehyde and benzoyl radicals, with the same rate reaction-constants as the corresponding reactions of methoxy radicals [27], [33].
- Reaction of benzaldehyde with NO producing benzoyl radicals and HNO. Reaction-rate constants have been defined similar to the analogous reaction of formaldehyde and NO [34].

# 4 DISCUSSION

Figure 1 shows that the proposed mechanism successfully predicts the consumption of the fuel and oxygen and the formation of products without any additive. Figure 2 compares the measured concentration profiles of n-heptane, ethylene and methanol with the profiles calculated using n-heptane models generated by the standard EA-version and by the modified EA-version. The changes made in the secondary mechanism for the oxidation of n-heptane allowed a significant improvement in the reproduction of species concentration profiles. Despite an overestimation of n-heptane consumption below 800K (Fig.2a), the amplitude of the reduction of fuel conversion in the NTC-region is better predicted by the modified EA-version. In addition, the calculated ethylene concentration profile (Fig.2b) is closer to the experimental observations and the methanol concentration peak around 650K is more accurately reproduced.

**FIGURE 2**

The following section first discusses the comparison between experimental and simulated results in the presence of NO, ethylene and methanol (Fig.1) and analyzes the reactions leading to the observed effects. The model is further used for predicting the impact of the addition of formaldehyde, ethane and CO on the fuel oxidation (Fig.3).

*4.1 Kinetic study of the influence of the addition of NO, ethylene and methanol*

*4.1.1  Addition of NO*

Figures 1a, 1b and 1c show that the complex impact of NO on the consumption of reactants is accurately predicted throughout the whole considered temperature range. The slight retarding effect of NO which is observed below 700K is mainly caused by reactions (1) of alkylperoxy radicals (ROO•) with NO to give alkoxy radicals:

$$ROO• + NO = RO• + NO_2 \quad (1)$$

These reactions compete with the isomerization of ROO• radicals followed by a second addition of oxygen yielding branching agents (hydroperoxides). The NO promoting effect observed above 700K is due to the reaction (2) between NO and •OOH radicals, transforming these rather unreactive radicals into •OH radicals. The following catalytic cycle can be written, with reaction (2) consuming NO and reactions (3-5) regenerating it:

$$NO + •OOH = NO_2 + •OH \quad (2)$$

$$NO_2 + •H = NO + •OH \quad (3)$$

$$NO_2 + •CH_3 = NO + CH_3O• \quad (4)$$

$$NO_2 + •C_2H_5 = NO + C_2H_5O• \quad (5)$$

The effect of this cycle is to noticeably reduce the concentration of •OOH radicals and to increase that of •OH radicals.

Figures 1d to 1k show that the effects of NO addition on the formation of the main products are correctly reproduced. In the main, these effects are:

➢ The important reduction in the formation of ethylene above 800K (Fig.1d) which is caused by reaction (5) becoming competitive with reaction (6), which is the main ethylene formation channel:

$$•C_2H_5 + O_2 = •OOH + C_2H_4 \quad (6)$$

➢ The decrease in the formation of methanol around 650K (Fig.1e) due to the NO consumption of methylperoxy radicals by reaction (7), which are the main source of methanol via reaction (8):

$$CH_3OO• + NO = CH_3O• + NO_2 \quad (7)$$

$$ROO\bullet + CH_3OO\bullet = RCHO + CH_3OH + O_2 \quad (8)$$

- The lower production of ethane (Fig.1f) which is caused by reaction (4) consuming $\bullet CH_3$ radicals and thus reducing the formation of ethane, obtained via reaction (9):

$$CH_3\bullet + CH_3\bullet = C_2H_6 \quad (9)$$

- The complex impact on the formation of formaldehyde above 700K (Fig.1g). As reactions (4) and (5) favour the formation of formaldehyde by an enhanced production of $CH_3O\bullet$ and $C_2H_5O\bullet$ radicals, the production of aldehydes increases due to the rapid decomposition of these radicals. This effect, enhanced by the increased global reactivity, explains the higher concentration of formaldehyde between 750 and 850K resulting from the addition of NO. At higher temperatures, the increased formation of $\bullet OH$ radicals promotes the consumption of formaldehyde and explains its lower concentration in presence of NO.

- The effects of NO on the concentrations of CO (Fig.1h), $H_2$ (Fig.1i), $H_2O$ (Fig.1j) and $CO_2$ (Fig.1k). The enhanced global reactivity induced by NO at temperatures above 650K results in an increased formation of CO and $H_2$. However, above 900K, the production of $\bullet OH$ radicals by reactions (2) and (3) becomes important and induces a larger consumption of CO thus producing a larger concentration of $CO_2$. As $\bullet OH$ radicals are consumed by formaldehyde producing $\bullet CHO$ radicals and $H_2O$, the production of water also increases. As the main source of $H_2$ is the H-abstractions by H-atoms from aldehydes, an increase of $\bullet OH$ production results in a $H_2$ concentration drop off (>900K) due to the competition between reactions (10) and (11):

$$HCHO + \bullet H = \bullet CHO + H_2 \quad (10)$$

$$HCHO + \bullet OH = \bullet CHO + H_2O \quad (11)$$

The favoured formation of $CO_2$ and water comes along with the increased consumption of oxygen above 900K when NO is added.

*4.1.2 Addition of ethylene and methanol*

Figures 1a and 1b show that the addition of ethylene and methanol results in a reduced global reactivity below 800K. This is correctly reproduced by the model. Ethylene and methanol are stable compounds and mainly consumed by the attack of •H and •OH radicals resulting in less reactive radicals and therefore representing a non negligible sink for reactive radicals. This competition in the consumption of •H and •OH radicals between the additives (ethylene, methanol) and the initial fuel reduces the oxidation-rate of n-heptane and toluene. At higher temperatures, this competition becomes negligible compared to the rate of production of •H and •OH radicals. Therefore, the impact of these additives on the fuel oxidation becomes much smaller.

For temperatures up to 1000K the production of ethylene, which is an important intermediate product of hydrocarbons oxidation, exceeds its consumption (Fig.1d). Only at higher temperatures the ethylene consumption is so strong that its concentration becomes quasi-independent of the initially added amount of ethylene.

At temperatures around 600K, the methanol concentration exceeds the amount of introduced methanol (Fig.1e). The observed concentration peak corresponds to the amount of methanol which is produced via reaction (8). At higher temperatures (>700K), the consumption of methanol exceeds its production and thus its concentration profile corresponds to that observed for the oxidation of n-heptane and toluene which is characterized by a NTC-zone between 700 and 800K. The presence of ethylene and methanol enhances the formation of formaldehyde through the depletion of •$C_2H_3$, •$CH_2OH$ and •$CH_3O$ radicals.

*4.3 Study of the effect of the addition of ethane, formaldehyde and CO*

The good agreement between experimental and modeling results justifies using the model for predicting the impact of various organic additives on the fuel oxidation. Ethane, formaldehyde and CO are known to be present in EGR and may influence the HCCI-engine control. The model was then used to study the chemical impact of these species on the fuel oxidation.

**FIGURE 4**

Figure 3a compares the impact of various amounts of formaldehyde on the oxidation of n-heptane. The simulations show that the addition of formaldehyde has a retarding effect on the n-heptane oxidation at low temperatures (<800K) but that it also increases the NTC-amplitude. At high concentrations of added formaldehyde (1600ppmv), the resulting retarding effect becomes so strong that up to 750K no reactivity of the mixture is predicted. Formaldehyde is consumed by the attack of •OH radicals via reaction (12), while the •CHO radicals react themselves by reaction (13):

$$HCHO + •OH = •CHO + H_2O \quad (12)$$

$$•CHO + O_2 = CO + •OOH \quad (13)$$

The sum of reactions (12) and (13) results in a net transformation of the reactive •OH into the less reactive •OOH radicals thus explaining the increased NTC-amplitude and the lack of reactivity at temperatures below 800K for high concentrations of added formaldehyde.

Figure 3b shows the impact of various amounts of ethane on the oxidation of n-heptane. The addition of ethane yields an inhibition at low temperatures, but also leads to a slight increase of the NTC-amplitude, which is less apparent than when formaldehyde is added. This amplification caused by ethane can be explained by the formation of ethyl radicals yielding an increased production of •OOH radicals via reaction (6).

Figure 3c compares the impact of various amounts of CO on the oxidation of n-heptane. The addition of CO slightly inhibits the n-heptane oxidation at low temperatures due to the consumption of •OH radicals via reaction (14), whereas at higher temperatures, the CO addition accelerates the consumption of n-heptane and the NTC-amplitude decreases via the •OH producing reaction (15):

$$CO + •OH = CO_2 + •H \quad (14)$$

$$CO + •OOH = CO_2 + •OH \quad (15)$$

These observations are in agreement with the analysis performed by Subramanian et. al. [4] on the impact of CO on ignition delay times.

## 5 CONCLUSION

Our study points out the influence of EGR compounds on the fuel oxidation in HCCI-engine applications. The experimental results and the developed kinetic model contribute to a deeper understanding of complex chemical impacts of EGR on the fuel oxidation control. Further work are planned by our group to also assess these effects on autoignition.

It confirms the observations made by Dubreuil [6] and Moréac [7], who have emphasized the importance of the impact of NO on the oxidation of hydrocarbons. Generally, the presence of NO lowers the overall fuel reactivity at low temperatures and increases it at higher temperatures. Therefore, the content of recycled NO might impact strongly the ignition delay in an HCCI-engine. The addition of NO reduces considerably the formation of ethylene, ethane, CO and $H_2$ above 900K and promotes those of CO and $H_2O$. Ethylene and methanol were experimentally and theoretically found to have little impact on the oxidation of n-heptane/toluene mixtures, even if a slight retarding effect on the fuel oxidation is observed at low temperatures. In contrast, a strong increase in the NTC-amplitude is predicted for the addition of formaldehyde. At high concentrations and at temperatures below 750K, its presence can completely inhibit the fuel oxidation. It was further shown that CO influences the oxidation kinetics of hydrocarbons and that therefore the impact of CO on HCCI-engine control might not only be limited to a diluting effect. Further experimental investigation is necessary since CO is among the main components of HCCI-engine EGR.

**REFERENCES**

[1]  X.C. Lü, W. Chen, Z. Huang, *Fuel* 84 (2005) 1974-1083.

[2]  X.C. Lü, W. Chen, Z. Huang, *Fuel* 84 (2005)1984-1092.

[3]  S.Tanaka, F. Ayala, J.C. Keck, J.B. Heywood, *Combust. Flame* 132 (2003) 219-239.

[4]  G. Subramanian, A. Pires da Cruz, R. Bounaceur, L. Vervisch, *Combust. Sci. Technol.* 179 (2007) 1937-1962.

[5]  T. Amano, F.L. Dryer, *27th Symposium of Combustion* (1998) 397-404.

[6]  A. Dubreuil, F. Foucher, C. Mounaim-Rousselle, G. Dayma, P. Dagaut, *Proc. Comb. Inst.* 31 (2007) 2879-2886.

[7]  G. Moréac, P. Dagaut, J.F. Roesler, M. Cathonnet, *Combust. Flame* 145 (2006) 512-520.

[8]  A. Piperel, X. Montagne, P. Dagaut, *SAE*, 2007-24-0087.

[9]  A.K. Agarwal, *Prog. Energ. Combust. Sci*. 33 (2007) 233-271.

[10]  P. Dagaut, M. Cathonnet, J.P. Rouan, R. Foulatier, A. Quilgars, J.C. Boettner, F. Gaillard, H. James, *J. Phys. E. Sci. Instrum*. 19 (1985) 207-209.

[11]  P. Dagaut, M. Reuillon, M. Cathonnet, *Combust. Flame* 101 (1994) 132-140.

[12]  P. Dagaut, A. Nicolle, *Combust. Flame* 114 (2005) 161-171.

[13]  F. Buda, R. Bounaceur, V. Warth, P.A. Glaude, R. Fournet, F. Battin-Leclerc, *Combust. Flame* 142 (2005) 170-186.

[14]  R. Bounaceur, I. da Costa, F. Fournet, *Int. J. Chem. Kin.* 37-1 (2005) 25-49.

[15]  C. Muller, V. Michel, G. Scacchi, G.M. Côme, *J. Chim. Phys.* 92 (1995) 1154-1178.

[16]  S.W Benson, *Thermochemical Kinetics, 2nd ed., John Wiley, New York* (1976).

[17]  R.J. Kee, F.M. Rupley, J.A Miller, Sandia Laboratories Report, *SAND 89 - 8009B* (1993).

[18]  V. Warth, N. Stef , P.A. Glaude, F. Battin-Leclerc, G. Scacchi, G.M. Côme, *Combust. Flame* 114 (1998) 81-102.

[19]  P. Barbé, F. Battin-Leclerc, G.M. Côme, *J. Chim. Phys.* 92 (1995) 1666-1692.

[20] D.L. Baulch, C.J. Cobos, R.A. Cox, P. Frank, G.D. Hayman, T. Just, J.A. Kerr, T.P. Murrells, M.J. Pilling, J. Troe, R.W. Walker, J. Warnatz, *Combust. Flame* 98 (1994) 59-79.

[21] W. Tsang, R.F. Hampson, *J. Phys. Chem. Ref. Data* 15 (1986) 1087.

[22] W. Tsang, *J. Phys. Chem. Ref. Data* 16 (1987) 471-508.

[23] J. Troe, *Ber. Buns. Phys. Chem.* 78 (1974) 478.

[24] P.D. Lightfoot, R.A. Cox, J.N. Crowley, M. Destriau, G.D. Hayman, M.E Jenkin, G. K. Moortgat, F. Zabel, *Atmos. Envir. A* 26 (1992) 1805-1961.

[25] J. Andrae, D. Johansson, P. Björnbom, P. Risberg, G. Kalghatgi, *Combust. Flame* 140 (2005) 267-286.

[26] M. Hori, Y. Matsunaga, N. Marinov, W.J. Pitz, C.K. Westbrook, *Proc. Combust. Inst.* 27 (1998) 389-396.

[27] G.P. Smith, D.M. Golden, M. Frenklach, N.W. Moriarty et al. in http://www.me.berkeley.edu/gri_mech/

[28] A.M. Dean, J.W. Bozzelli, *Combustion Chemistry, Gardiner, W., Jr., Ed., Springer-Verlag: New York* (1997).

[29] R. Atkinson, D.L. Baulch, R.A. Cox, R.F. Hampson, Jr. Kerr, J. Troe, *J. Phys. Chem. Ref. Data* 21 (1992) 1125.

[30] N. Marinov, *N. M. LLNL Report No. UCRL-JC-129372, Lawrence, Livermore National Laboratories, Berkeley, CA* (1998).

[31] P.A. Glaude, N. Marinov, Y. Koshiishi, N. Matsunaga, M. Hori, *Energy Fuels* 19 (2005) 1839-1849.

[32] N.K. Srinivasan, M.C. Su, J.W. Sutherland, J.V. Michael, *J. Phys. Chem. A* 109 (2005) 1857-1863.

[33] R. Atkinson, D.L. Baulch, R.A. Cox, J.N. Crowley, R.F. Hampson, R.G. Hynes, M.E. Jenkins, M.J. Rossi, J. Troe, Atmos. *Chem. Phys. Discuss.* 5 (2005) 6295-7168.

[34] W. Tsang, J.T. Herron, *J. Phys. Chem. Ref. Data* 20 (1991) 609-663.

**FIGURES**

**Figure 1**:

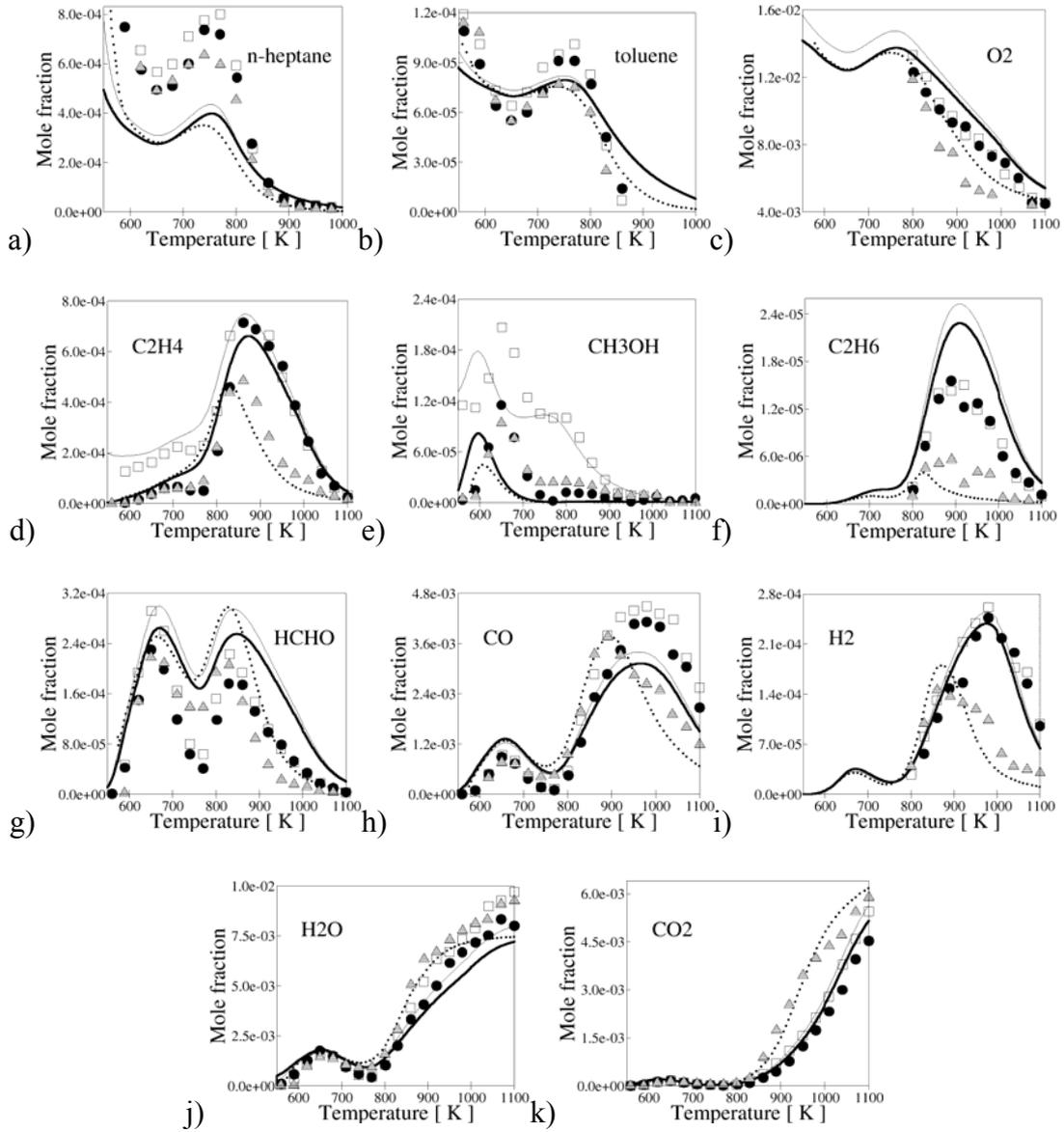
18

**Figure 2**:

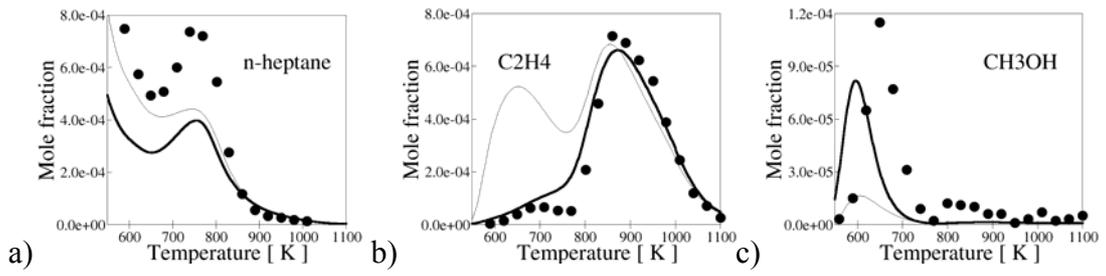

a) b) c)

**Figure 3**:

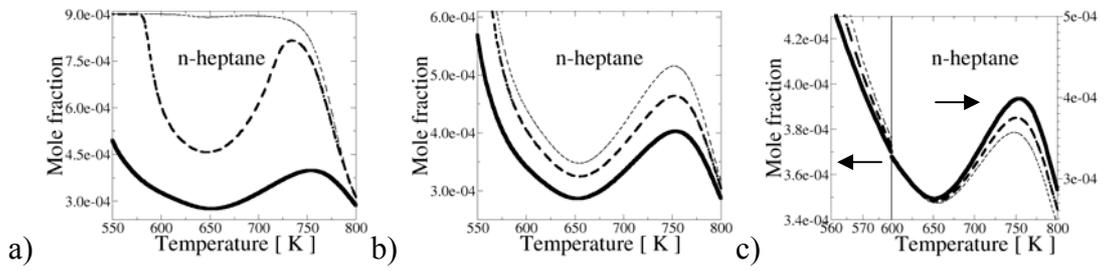

a) b) c)

**FIGURE CAPTIONS**

**Figure 1**: Experimental (symbols) and simulated (lines) mole fractions of (a) n-heptane, (b) toluene, (c) oxygen and (d-k) the main combustion intermediates with no additive (black circles ● and thick full lines ▬), with addition of ethylene and methanol (white squares ☐ and thin lines ─) and with addition of NO (grey shaded triangles ∆ and dotted lines ---).

**Figure 2**: Experimental (symbols) and simulated (lines) mole fractions of (a) n-heptane, (b) ethylene and (c) methanol without additive. Simulations were performed with the modified EA-version (thick full lines ▬) and with the standard EA-version (thin full lines ─) mechanisms.

**Figure 3**: Simulated mole fractions of n-heptane with (a) formaldehyde-additive, (b) ethane-additive and c) CO-additive. Additive mole fractions are 0ppmv (thick line ▬), 800ppmv (medium dotted line ---) and 1600ppmv (thin dotted line ---).

**LIST OF SUPPLEMENTAL MATERIAL :**

1 supplemental file:   KINETIC-MECHANISM.S1.txt

**CAPTION FOR SUPPLEMENTAL MATERIAL :**

Kinetic reaction mechanism for the oxidation of n-heptane-toluene blends, valid for the addition of $C_1$-$C_2$ carbon species and NO written in the CHEMKIN format.